\documentclass[superscriptaddress,twocolumn,prb]{revtex4}
\usepackage[latin1]{inputenc}
\usepackage{amsmath}
\usepackage{amsfonts}
\usepackage{amssymb}
\usepackage{braket}
\usepackage{tikz}

\begin{document}
\title{High-accuracy ab-initio quantum chemistry by means of an SU(2) $\times$ U(1) invariant matrix product state Ansatz: the static second hyperpolarizability}
\author{Sebastian Wouters}
\email{sebastian.wouters [at] ugent.be}
\affiliation{Center For Molecular Modeling, Ghent University, Ghent, Belgium}
\author{Peter A. Limacher}
\affiliation{Department of Chemistry, McMaster University, Hamilton, Ontario, Canada}
\author{Dimitri Van Neck}
\affiliation{Center For Molecular Modeling, Ghent University, Ghent, Belgium}
\author{Paul W. Ayers}
\affiliation{Department of Chemistry, McMaster University, Hamilton, Ontario, Canada}

\begin{abstract}
{\centering \textbf{Poster presentation at the 16th International Conference on Recent Progress in Many-Body Theories, Nov. 28 to Dec. 2, 2011, Bariloche, Argentina.}}\\

We have implemented the single-site density matrix renormalization group algorithm for the variational optimization of SU(2) $\times$ U(1) (spin and particle number) invariant matrix product states for general spin and particle number symmetric fermionic Hamiltonians. This class also includes non-relativistic quantum chemical systems within the Born-Oppenheimer approximation. High-accuracy ab-initio finite field results of the longitudinal static polarizabilities and second hyperpolarizabilities of one-dimensional hydrogen chains are obtained with the algorithm. A comparison with other methods is made.
\end{abstract}

\maketitle

\section{Introduction}
The density matrix renormalization group (DMRG) outperforms other high-accuracy methods for one-dimensional systems. The reason is its underlying matrix product state (MPS) Ansatz, which efficiently captures the quantum entanglement properties of low-lying eigenstates. This allows to reach exact diagonalization accuracy within a reasonable amount of time. Sections \ref{sectMPS} and \ref{sectDMRG} describe the MPS Ansatz and the DMRG algorithm. Section \ref{sectionsymmetry} discusses how global SU(2) $\times$ U(1) symmetry is imposed on an MPS and how the SU(2) $\times$ U(1) symmetry of the fermionic Hamiltonian allows to work with reduced tensors only. The finite field method is explained in section \ref{sectFF}. The results are shown in section \ref{sectRES} and analyzed in section \ref{sectDIS}. Finally, a conclusion is made in section \ref{sectCON}. Note that this paper corresponds to a poster presentation at the 16th International Conference on Recent Progress in Many-Body Theories and is therefore by no means complete in terms of references.
\section{Matrix product states\label{sectMPS}}
A quantum mechanical many-body system consists of a Hamiltonian acting on a Hilbert space. The Hilbert space is formed as the direct product of $L$ single particle spaces. Its dimension increases exponentially with $L$. This has induced the development of many approximative solution methods. DMRG is one of them. It can be understood in two ways: as a renormalization group technique or by means of its underlying Ansatz, the MPS. This text focuses on the latter.\\

In \cite{schollwockDMRGatMPSage} is shown that every quantum state can be rewritten as an MPS:
\begin{equation}
\ket{\Psi} = \sum\limits_{\{ \alpha_j \}} \sum\limits_{\{i_j\}} A^{\alpha_1}_{i_1} A^{\alpha_2}_{i_1 i_2} ... A^{\alpha_{L-1}}_{i_{L-2} i_{L-1}} A^{\alpha_L}_{i_{L-1}} \ket{\alpha_1 ... \alpha_L}.
\end{equation}
The indices $\alpha_j$ represent the local (single particle) degrees of freedom. The indices $i_j$ are called the virtual indices. For $\ket{\Psi}$ to represent the full Hilbert space, the dimension of the virtual indices has to grow exponentially toward the middle of the MPS chain. For practical calculations, these virtual dimensions are truncated to $D$ and the MPS forms a variational Ansatz for the given problem. For one-dimensional systems, this Ansatz efficiently captures the quantum entanglement properties of low-lying eigenstates.
\section{The density matrix renormalization group\label{sectDMRG}}
A general ab-initio quantum chemistry Hamiltonian is written in second-quantized form as:
\begin{eqnarray}
\hat{H} & = & \sum\limits_{\alpha, \beta, \sigma} (\alpha \mid \hat{T} \mid \beta) \hat{a}^{\dagger}_{\alpha \sigma} \hat{a}_{\beta \sigma} \nonumber\\ & + & \frac{1}{2} \sum\limits_{\alpha,\beta,\gamma,\delta, \sigma, \tau} (\alpha \beta \mid \hat{V} \mid \gamma \delta) \hat{a}^{\dagger}_{\alpha \sigma} \hat{a}^{\dagger}_{\beta \tau} \hat{a}_{\delta \tau} \hat{a}_{\gamma \sigma}
\end{eqnarray}
where $\alpha$, $\beta$, $\gamma$ and $\delta$ denote orbitals and $\sigma$ and $\tau$ denote spin projections. The global spin and the global particle number are conserved by this Hamiltonian.\\
              
\begin{figure}[t]
 \includegraphics[width=0.38\textwidth]{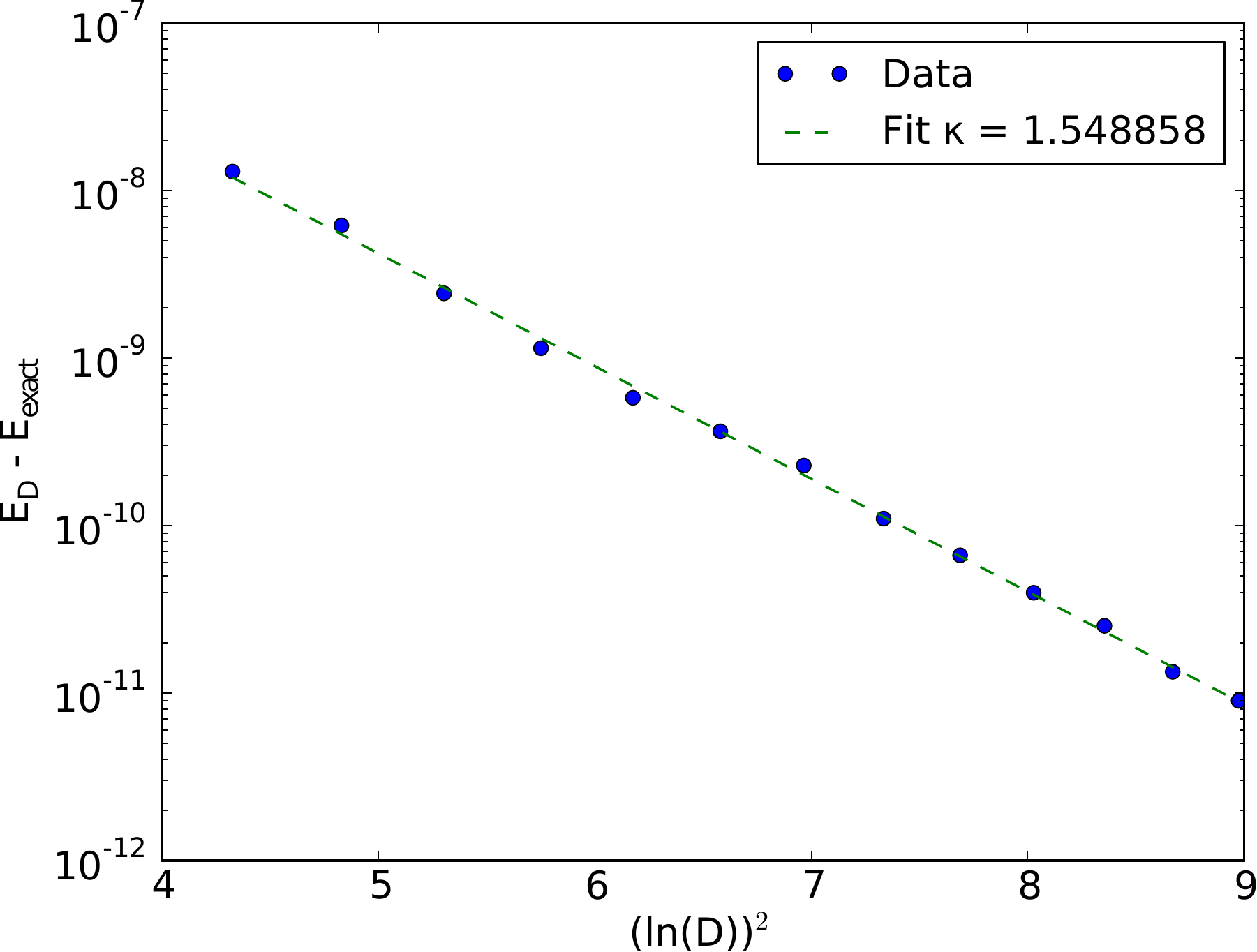}
 \caption{\label{scalingplot} The ground state of a hydrogen chain of 36 atoms, with an alternate atom spacing of 2/3 a.u., in the L\"owdin transformed STO-6G basis, is approximated by several MPS Ans\"atze. The scaling of the ground state energy with $D$, the virtual dimension per symmetry sector (see section \ref{sectionsymmetry}), follows \eqref{chanlaw}.}
\end{figure}

The single-site DMRG algorithm sweeps through the MPS chain and optimizes at each step a single $A$-tensor variationally, while the others are kept constant. When no local minima are encountered, the global minimum of $\braket{\Psi \mid \hat{H} \mid \Psi}$ is found and the whole MPS is variationally optimized. The size of the virtual dimension controls the accuracy of the result \cite{2002JChPh.116.4462C}:
\begin{equation}
\ln(E_D - E_{\mathrm{exact}}) = a - \kappa \left( \ln(D) \right)^2 \label{chanlaw}
\end{equation}
Here, $a$ and $\kappa$ are fitting parameters, $E_{\mathrm{exact}}$ is the exact diagonalization result and $E_D$ the energy when an Ansatz with virtual dimension $D$ is used. The relation \eqref{chanlaw} is illustrated in Fig. \ref{scalingplot}.
\section{Imposing global SU(2) $\times$ U(1) symmetry \label{sectionsymmetry}}
The use of global symmetries of the Hamiltonian has two main advantages \cite{2002EL57852M}:
\begin{itemize}
 \item It enables to scan only the region of interest of the many-body Hilbert space.
 \item It causes an improvement in algorithmic performance: CPU and memory demands decrease.
\end{itemize}
              
Global symmetry can be imposed by requiring that the three-index tensors $A^{\alpha}_{L R}$ in the MPS chain are irreducible tensor operators of the imposed symmetry group \cite{2002EL57852M, PhysRevA.82.050301}. The local and virtual bases are represented in states with the correct symmetry, i.e. spin $s$ or $j$, spin projection $s^z$ or $j^z$ and particle number $N$. Due to the Wigner-Eckart theorem, each $A$-tensor decomposes in a structural part and a degeneracy part $T$:
\begin{eqnarray}
 A^{\alpha}_{L R} & = & A^{(s s^z N)}_{(j_L j^z_L N_L \beta_L)(j_R j^z_R N_R \beta_R)} \\ & = & \braket{ j_L j^z_L s s^z \mid j_R j^z_R } \delta_{N_L + N, N_R} T^{(s N)}_{(j_L N_L \beta_L)(j_R N_R \beta_R)} ~ ~ ~
\end{eqnarray}
$\beta$ counts the number of times the irreducible representation $(j N)$ occurs at a virtual boundary \footnote{In DMRG calculations with smaller global symmetry, basis states with a certain symmetry are sometimes explicitly kept in the reduced basis to avoid losing quantum numbers \cite{2004JChPh.120.3172C}. Note that the division of the virtual indices in the symmetry sectors $(j N)$ does exactly the same.}. If $D(j N) = \text{size}(\beta)$, this corresponds to a virtual dimension of $(2j+1)D(jN)$ in a non-symmetry adapted MPS. These virtual dimensions are truncated per symmetry sector to $D$. The desired global symmetry can be imposed by requiring that the left virtual index of the leftmost tensor in the MPS chain consists of 1 irreducible representation corresponding to $(j_L N_L) = (0 0)$, while the right virtual index of the rightmost tensor consists of 1 irreducible representation corresponding to $(j_R N_R) = (S N)$, the desired global spin and particle number.\\
              
The operators $\hat{b}^{\dagger}_m = \hat{a}^{\dagger}_m$ and $\hat{b}_m = (-1)^{\frac{1}{2}-m}\hat{a}_{-m}$ transform as irreducible tensor operators with spin $\frac{1}{2}$ under SU(2). All (partial) evaluations of creation/annihilation operators can hence be done by implicitly summing over the common multiplets and recoupling the local, virtual and operator spins. No spin projections or Clebsch-Gordan coefficients are used in the algorithm \footnote{To the best of our knowledge, full SU(2) symmetry has been implemented only once in ab-initio quantum chemistry DMRG calculations \cite{2008JChPh.128a4107Z}. However, in \cite{2008JChPh.128a4107Z} no use is made of the Wigner-Eckart theorem to work with reduced tensors.}.
\section{Finite field extrapolations\label{sectFF}}
\begin{figure*}[t]
 \includegraphics[width=0.38\textwidth]{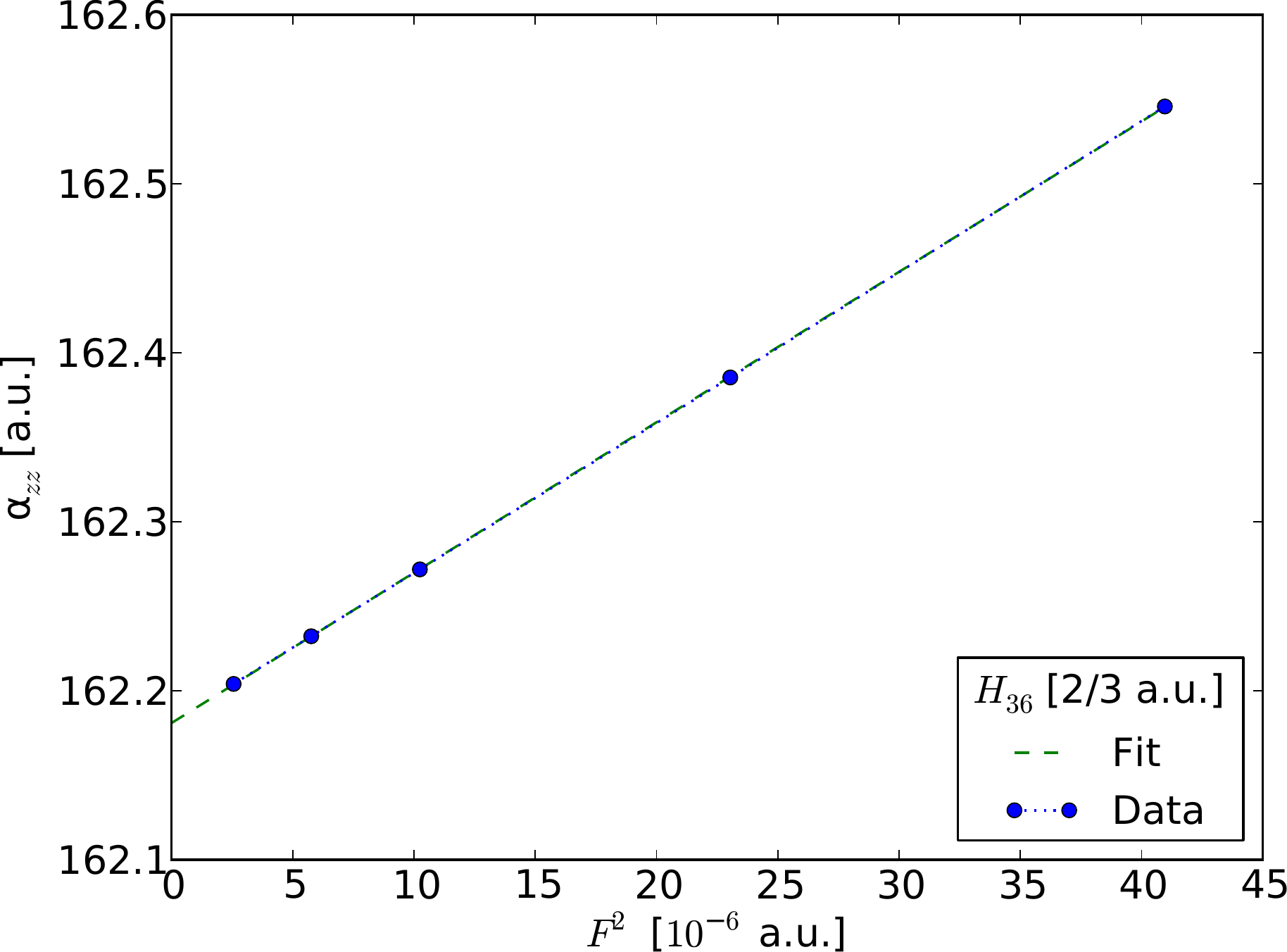} \hspace{0.05\textwidth} \includegraphics[width=0.38\textwidth]{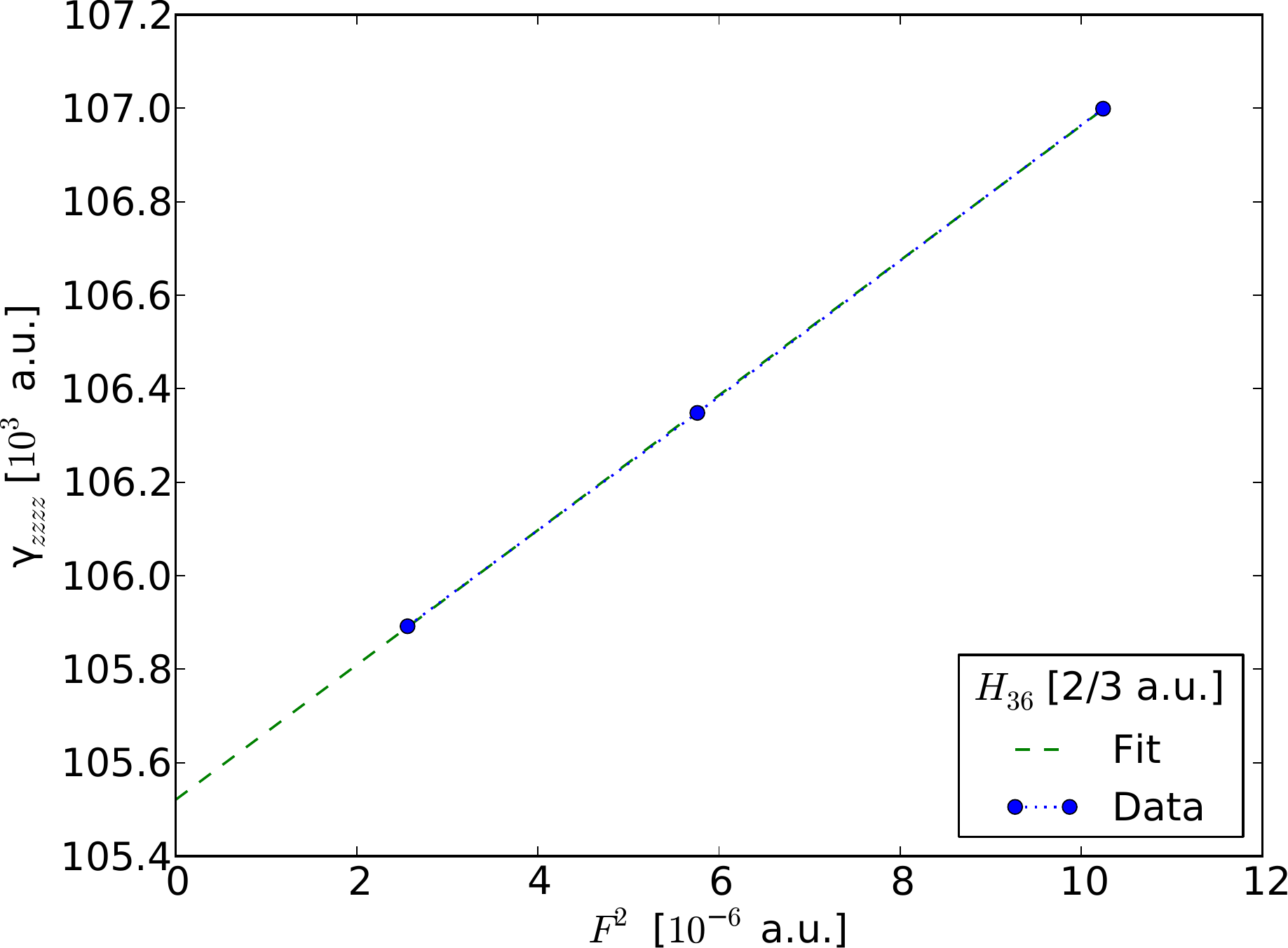}
 \caption{\label{extrapolplot} Finite field extrapolations for MPS calculations of a hydrogen chain of 36 atoms, with an alternate atom spacing of 2/3 a.u., in the L\"owdin transformed STO-6G basis.}
\end{figure*}

The longitudinal static polarizability $\alpha_{zz}$ and second hyperpolarizability $\gamma_{zzzz}$ can be obtained for centrosymmetric systems as \cite{2009IJQC..109.3103C}:
\begin{eqnarray}
 \alpha_{zz} & = & \left(\frac{2E(0) -2E(F)}{F^2}\right)_{F \rightarrow 0}\\
 \gamma_{zzzz} & = & \left(\frac{-6E(0) +8E(F) -2 E(2 F)}{F^4}\right)_{F \rightarrow 0}
\end{eqnarray}
where $E(F)$ denotes the energy when an electric field $F$ in the $z$-direction is applied. If $F$ is chosen too small, noise due to finite precision arithmetic is enhanced. If $F$ is chosen too large, higher order effects come into play. The minimal finite difference formulas both have a lowest order correction term of $\mathcal{O}(F^2)$. We calculate both $\alpha_{zz}$ and $\gamma_{zzzz}$ for several field values $F$ and extrapolate these to $F = 0$ with
\begin{equation}
q(F) = q(0) + c F^2
\end{equation}
where $q$ can be $\alpha_{zz}$ or $\gamma_{zzzz}$. Values of $q(0)$ and $c$ are obtained by the fit. The procedure is shown in Fig. \ref{extrapolplot}.

\section{Results\label{sectRES}}
\begin{figure*}[t]
 \includegraphics[width=0.38\textwidth]{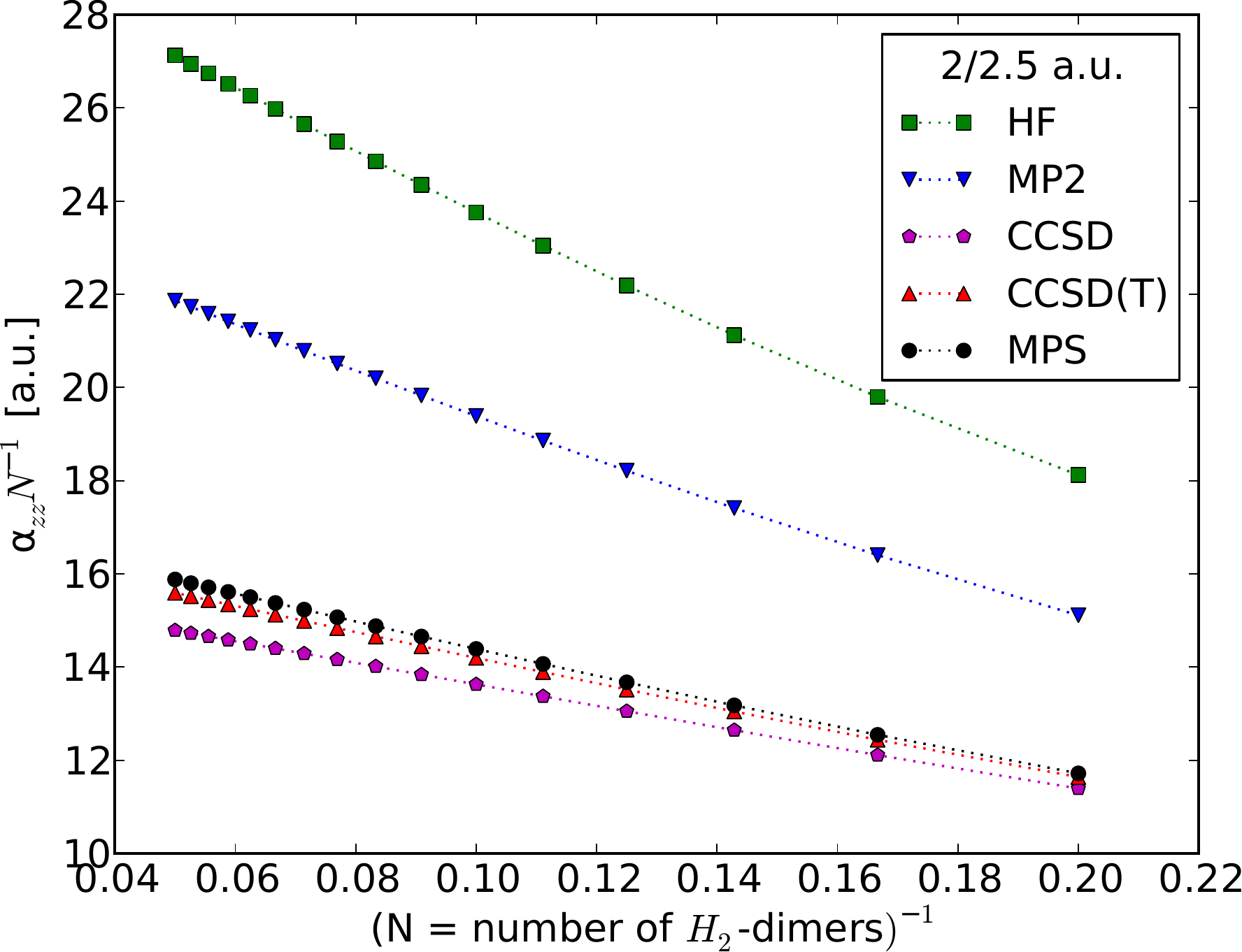} \hspace{0.05\textwidth} \includegraphics[width=0.38\textwidth]{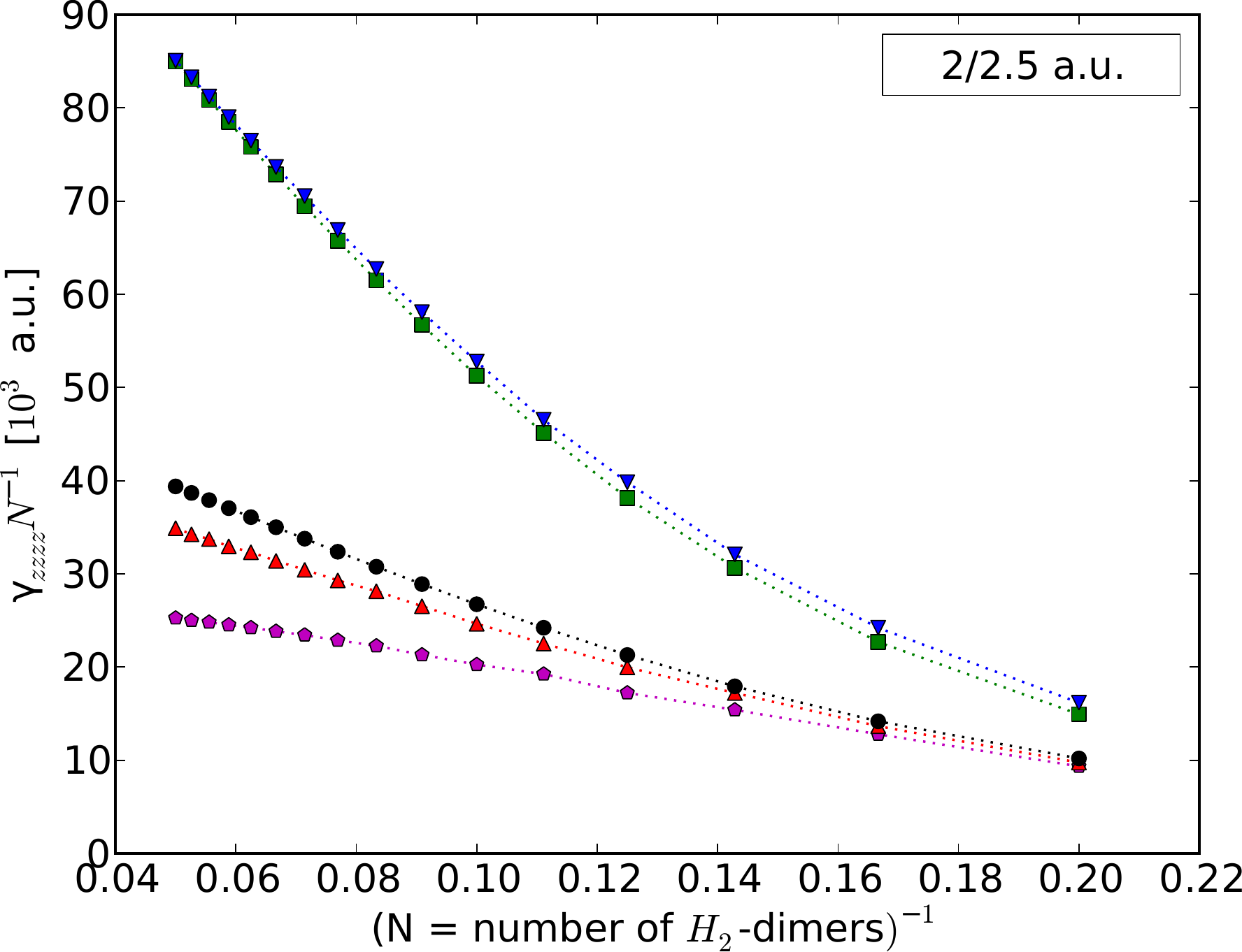}
 \caption{\label{result1} Polarizabilities and second hyperpolarizabilities of hydrogen chains with an alternate atom spacing of 2/2.5 a.u., in the L\"owdin transformed STO-6G basis, for different levels of theory.}
\end{figure*}

\begin{figure*}[t]
 \includegraphics[width=0.38\textwidth]{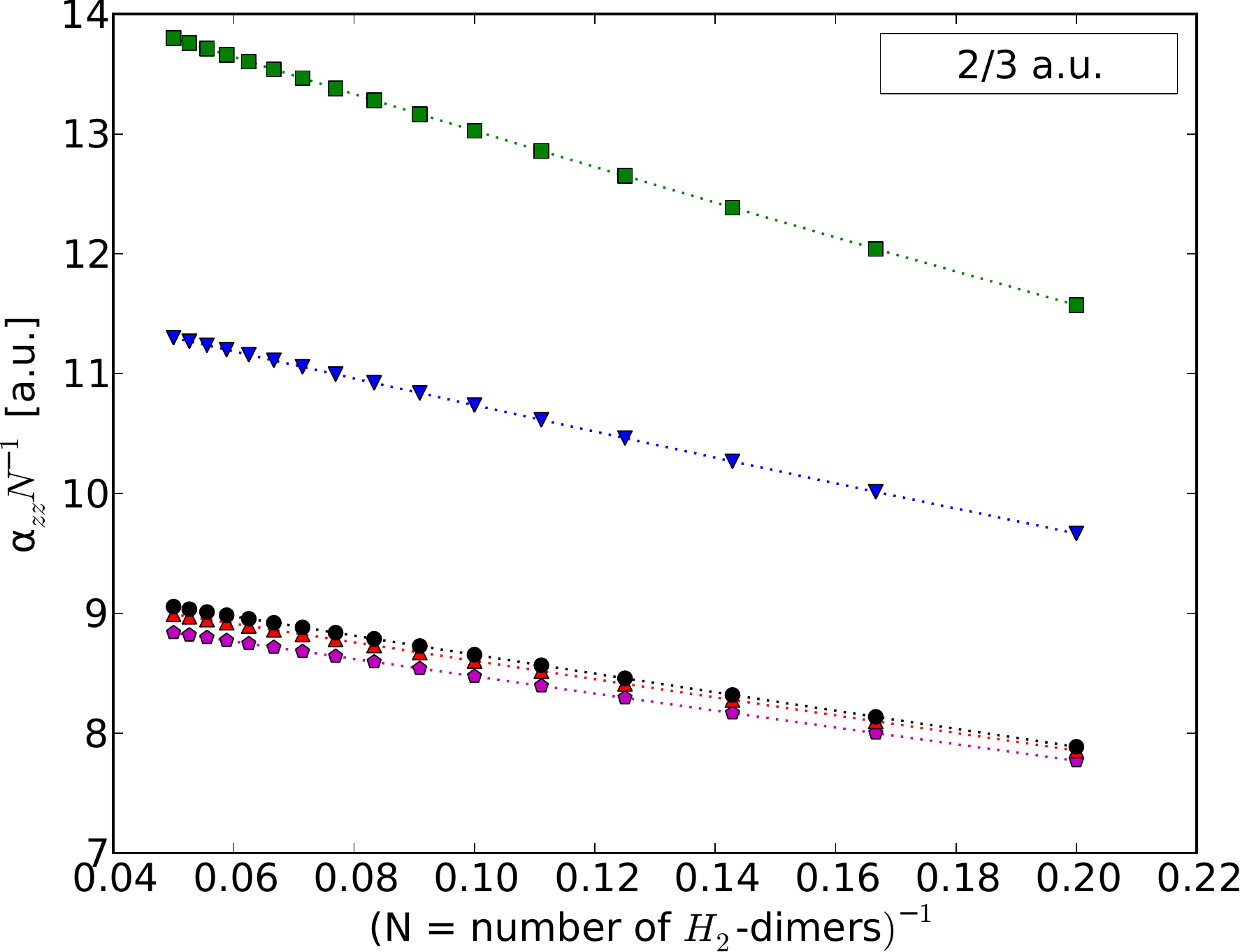} \hspace{0.05\textwidth} \includegraphics[width=0.38\textwidth]{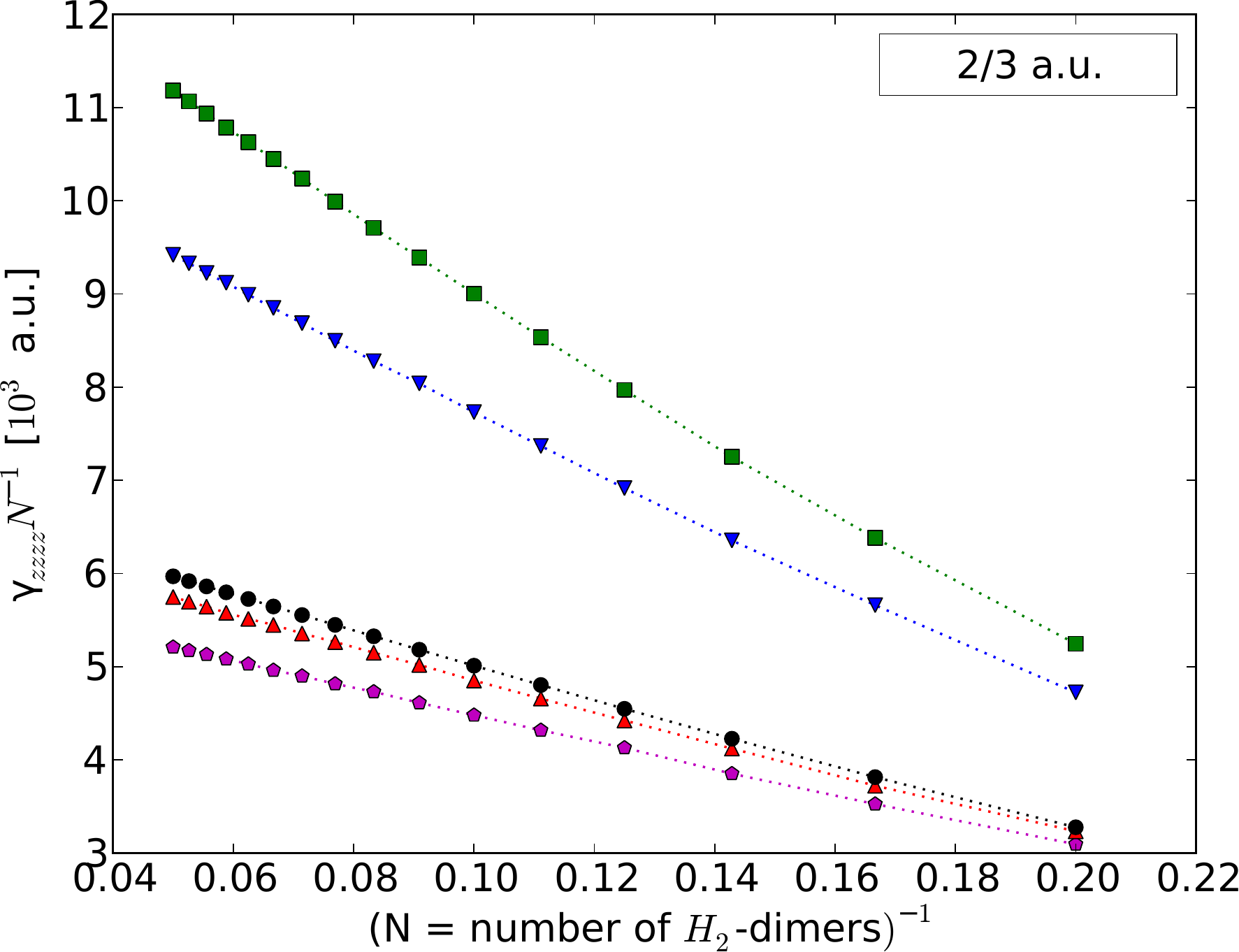}
 \caption{\label{result2} Polarizabilities and second hyperpolarizabilities of hydrogen chains with an alternate atom spacing of 2/3 a.u., in the L\"owdin transformed STO-6G basis, for different levels of theory. The legend is shown in Fig. \ref{result1}.}
\end{figure*}

\begin{figure*}[t]
 \includegraphics[width=0.38\textwidth]{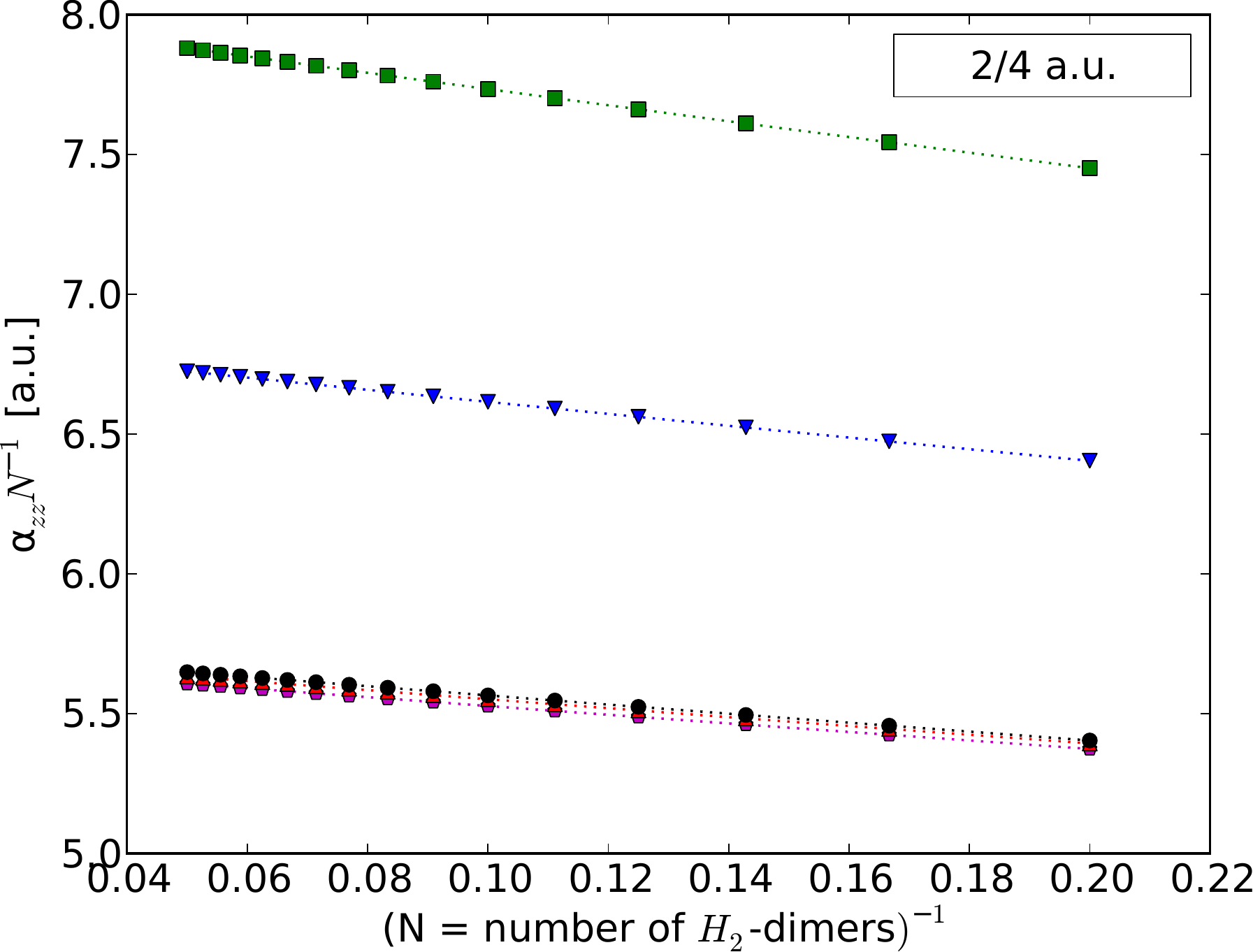} \hspace{0.05\textwidth} \includegraphics[width=0.38\textwidth]{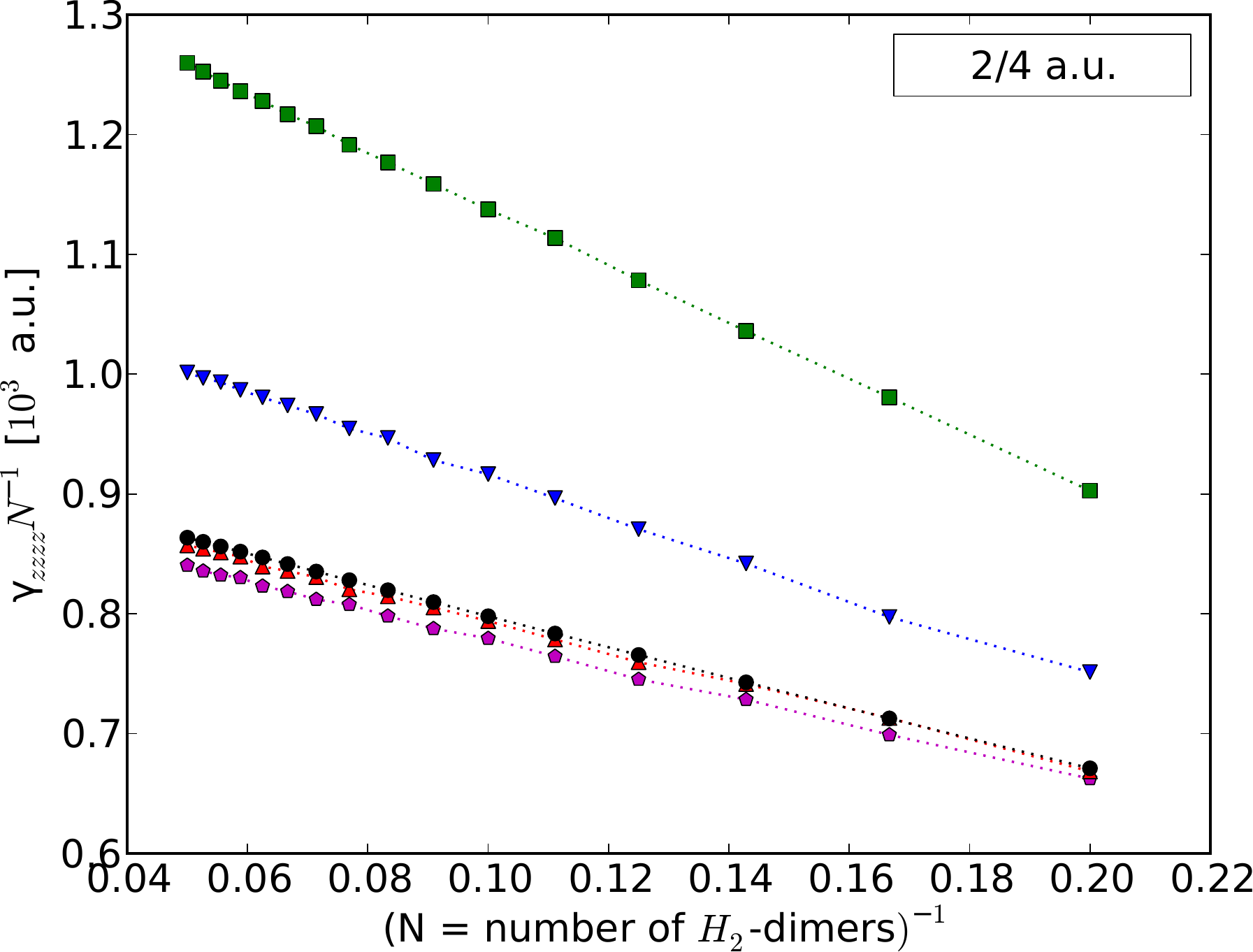}
 \caption{\label{result3} Polarizabilities and second hyperpolarizabilities of hydrogen chains with an alternate atom spacing of 2/4 a.u., in the L\"owdin transformed STO-6G basis, for different levels of theory. The legend is shown in Fig. \ref{result1}.}
\end{figure*}

The finite field extrapolation was carried out for hydrogen chains with
\begin{itemize}
 \item increasing length: $H_{2N}$ with $5 \leq N \leq 20$
 \item different bond length alternations: 2/2.5 a.u., 2/3 a.u. and 2/4 a.u.
 \item the L\"owdin transformed minimal basis set STO-6G
 \item 5 levels of theory: HF, MP2, CCSD, CCSD(T) and MPS
\end{itemize}
The virtual dimension $D$ of the symmetry sectors of the MPS was chosen in order that the MPS results match the exact diagonalization results. HF, MP2, CCSD and CCSD(T) calculations were performed with the molecular electronic structure program Dalton\cite{daltonref}. The extrapolation results are shown in Fig. \ref{result1}, \ref{result2} and \ref{result3}.
\section{Discussion\label{sectDIS}}
The change in bond length alternation results in a different separation of $H_2$ molecules, which themselves have a fixed bond length of 2 a.u. The multireference character doesn't change significantly for the three bond length alternations, but the correlation length of the electrons in the lattice does.\\

The results for $\alpha_{zz}$ and $\gamma_{zzzz}$ are ordered in magnitude as:
\begin{equation}
	CCSD < CCSD(T) < MPS < MP2 < HF
\end{equation}

This ordering differs from \cite{2009IJQC..109.3103C}, where an aug-cc-pVDZ basis set was used for the HF, MP2, CCSD and CCSD(T) methods. The ordering is hence basis set dependent. The results for $\alpha_{zz}$ and $\gamma_{zzzz}$ are ordered according to their deviation from the exact diagonalization results as:
\begin{equation}
HF > MP2 > CCSD > CCSD(T) > MPS
\end{equation}
which confirms that CCSD(T) is more accurate to calculate the polarizability and second hyperpolarizability than HF, MP2 and CCSD \cite{2009IJQC..109.3103C}.\\

The deviation between the CCSD(T) and exact diagonalization results for $\alpha_{zz}$ and $\gamma_{zzzz}$ is the largest for the 2/2.5 a.u. bond length alternation. The electrons in the 2/2.5 a.u. configuration are more mobile throughout the chain than the electrons in the 2/4 a.u. configuration. They are thus correlated on a longer length scale in the lattice, i.e. with more electrons. Results obtained with a method that captures correlation only partially are expected to deviate more from the true results in that case. Fortunately, the amount of electron correlation captured by the MPS Ansatz is controlled by the truncation of the virtual dimension and can be finetuned to any desired accuracy.
\section{Conclusion\label{sectCON}}
The MPS Ansatz is able to capture all required correlation to be able to obtain highly accurate ab-initio finite field results of the longitudinal static polarizability and second hyperpolarizability of one-dimensional chains.
\begin{acknowledgments}
This research was supported by Fonds Wetenschappelijk Onderzoek (FWO) and was carried out using the Stevin Supercomputer Infrastructure at Ghent University, funded by Ghent University, the Hercules Foundation and the Flemish Government - department EWI.
\end{acknowledgments}

\bibliographystyle{phjcp}
\bibliography{biblio}

\end{document}